\documentclass{article}

\usepackage{authblk}
\usepackage[numbers]{natbib}
\usepackage{amsfonts}
\usepackage{amsthm}
\usepackage{relsize}
\usepackage{amsmath}
\usepackage{geometry}
\usepackage{lscape}
\usepackage{ctable}
\usepackage{arydshln}
\usepackage[hidelinks]{hyperref}
\usepackage[shortlabels]{enumitem}
\usepackage{extarrows}
\usepackage[normalem]{ulem}
\usepackage{mdframed}
\usepackage{float}
\usepackage{tcolorbox}
\usepackage{pbox}
\usepackage{xcolor}

\title{Linearly representable games and pseudo-polynomial calculation of the Shapley value}
\author{Ferenc Illés\thanks{ferenc.illes@uni-corvinus.hu}}
\affil{Department of Finance, Corvinus University of Budapest}

\newtheorem{thm}{Theorem}[section]
\newtheorem{defi}{Definition}[section]

\newtheorem{lemma}{Lemma}[section]
\theoremstyle{definition}
\newtheorem{rem}{Remark}[section]
\newtheorem{ex}{Example}[section]
\newtheorem{cor}{Corollary}[section]

\definecolor{darkblue}{RGB}{0, 102, 0}
\definecolor{darkred}{RGB}{153, 0, 0}

\newcounter{note}
\newenvironment{note}{
\stepcounter{note}
\begin{equation*}
\tag{N\thenote}
}
{\end{equation*}\ignorespacesafterend}

\newcounter{sh}
\newenvironment{sh}{
\stepcounter{sh}
\begin{equation*}
\tag{S\thesh}
}
{\end{equation*}\ignorespacesafterend}

\floatstyle{plain}
\newfloat{alg}{h}{alg}[section]
\floatname{alg}{Algorithm}

\begin{document}
\maketitle

\begin{abstract}
    We introduce the notion of linearly representable games. 
    Broadly speaking, these are TU games that can be described by as many parameters as the number of players, 
    like weighted voting games, airport games, or bankruptcy games.  
    We show that the Shapley value calculation is pseudo-polynomial for linearly representable games. 
    This is a generalization of many classical and recent results in the literature (\citet{mann1962values}, \citet{AZIZ2013499}). 
    Our method naturally turns into a polynomial algorithm when the parameters are polynomial in the number of players.\\    
    
\end{abstract}

\section{Introduction}
The calculation of the Shapley value for games with transferable utility (TU) is a challenging task since its introduction in Shapley's famous paper \cite{shapley1953value}. Though most games can be described by a few parameters, the Shapley value is the mean of exponentially many numbers, which cannot be calculated in practice based on its definition if the number of players is large. In the recent 70 years many great algorithms have been developed to overcome the difficulties. It turned out that TU games are diverse in the complexity of the Shapley value, it is polynomial for some game classes (see for example \citet{dori1}, \citet{CostAlloc} and \citet{Airport}), but it can also be \mbox{NP-hard} (see \citet{complex} or \citet{complex2}). 
We have no good algorithms to solve \mbox{NP-hard} problems, and we probably never will. There is, however, a compromise. Some \mbox{NP-complete} problems can be solved in pseudo-polynomial time. An algorithm is pseudo-polynomial if its runnig time is a polynomial function of the \textit{numeric value} of the input. This is different from a polynomial algorithm, where the running time is a polynomial function of the \textit{size} of the input. The difference is crucial, because inputs of computational problems are usually numbers and numbers are represented by their digits and the number of digits of a number (roughly) equals its logarithm. Therefore, a pseudo-polynomial algorithm can still be exponential in the worst case, but completely usable in practice if the parameters are moderate.

The first pseudo-polynomial algorithm for the Shapley value was introduced by \citet{mann1962values} for weighted voting games, while \citet{complex} showed that the Shapley value calculation is NP-hard for this game class. Since then, similar results have been proved for other game classes, like bankruptcy games by \citet{AZIZ2013499} or liability games by \citet{CSOKA2021}. On the other hand, \citet{spanningTreeHard} proved that the Shapley value calculation for minimum spanning tree games is NP-hard even if the value of each coalition is $0$ or $1$. This obviously implies that there cannot be a pseudo-polynomial algorithm for such games as it would automatically be polynomial because of the bounded weights\footnote{In other words this problem is strongly NP-hard.}.

The crucial difference between the bankruptcy or liability games and the minimum spanning tree games is that the former ones can be described by assigning a number (a weight) to each player such that the value of each coalition is a function of the sum of the weights of its members, while minimum spanning tree games can be described by a weighted graph, which is a ``more complicated" representation.
We abstract this common property of bankruptcy games, liability games, weighted voting games and many other game classes by introducing the notion of \textit{linearly representable} games. It turns out the abovementioned positive result, the existance of a pseudo-polynomial algorithm for the Shapley value, holds for every linearly representable game. The algorithm can also be slightly modified to calculate the Shapley value for some special game classes, for example airport games\footnote{It is known that the Shapley value for airport games is polynomial, but we generalize this result as well.} in polynomial time and space.

The rest of the paper is structured as follows. In Section 2 and 3 we introduce the basic notations and the concept of linearly representable games, respectively.
In Section 4, we provide our main result, a pseudo-polynomial algorithm to calculate the Shapley value of one player in a linearly representable TU game and in Section 5 we discuss how the algorithm can be modified to get the Shapley values of all of the $n$ players without repeating the whole calculation $n$ times.
In Section 6 we broadly discuss possible extensions, special cases, and generalizations.

\section{Preliminaries}
A cooperative game with transferable utility is a pair $(N,v)$, where $N$ is a finite set of players and the characteristic function $v:2^N\mapsto \mathbb{R}$ is a function over the subsets of players such that $v(\emptyset)=0$. The number of players is denoted by $n=|N|$. Subsets of $N$ are called coalitions. The Shapley value assigns a unique value to each player as follows. 
\begin{itemize}
\item Let $i \in N$ be a player and let $S\subseteq N$ be a coalition such that $i\notin S$. The marginal contribution of player $i$ to coalition $S$ is $v'_i(S) = v(S\cup\{i\}) - v(S)$.
\item Let $\pi$ be a permutation of $N$. Let $Pre_\pi(i)$ denote the set of players preceding $i$ in $\pi$.
\item The Shapley value of player $i$ is the mean of the marginal contributions accross all possible $n!$ permutations of the players:
\begin{sh}
Sh(i)=\frac{1}{n!}\sum \limits_{\pi \in \Pi_N} v'_i(Pre_\pi(i)).
\label{sh::1}
\end{sh}
where $\Pi_N$ stands for the set of permutations over $N$. 
\end{itemize}

It is clear that any two terms in Formula \eqref{sh::1} are equal if the sets of players preceding player $i$ are the same: $v'_{i}(\pi_1)=v'_{i}(\pi_2)$ if $Pre_{\pi_1}(i)=Pre_{\pi_2}(i)$. 
If we make the summation over these equivalent terms, we get the formula
\begin{sh}
Sh(i)=\sum \limits_{S \subseteq N \setminus \{i\} } \frac{|S|!(n-|S|-1)!}{n!} \cdot v'_i(S) 
= \sum \limits_{S \subseteq N_{-i} } \frac{|S|!(n-|S|-1)!}{n!} \cdot v'_i(S),
\label{sh::2}
\end{sh}
where $N_{-i} = N\setminus\{i\}$.
The number of terms in expression (\ref{sh::2}) is $2^{n-1}$, which is definitely much less then $n!$, the number of terms in Formula (\ref{sh::1}), but this does not help much to calculate it, since it is still exponential.

\subsection{Integral representation}
Let us consider the Euler integral
\[\beta(a,b) = \int \limits_0^1 x^{a-1}(1-x)^{b-1} \ dx,\]
where $a$ and $b$ are nonnegative integers. With integration by parts we get that
\[\int \limits_0^1 x^{a-1}(1-x)^{b-1} \ dx = \underbrace{\left[\frac{x^a}{a}(1-x)^{b-1}\right]_{x=0}^{x=1}}_0-\int \limits_0^1 -\frac{x^a}{a}(b-1)(1-x)^{b-2} \ dx, \]
which means that $\beta(a,b) = \frac{b-1}{a}\beta(a+1,b-1)$.
By induction on $b$ we can obtain the well-known formula \[\beta(a,b) = \frac{(a-1)!\cdot (b-1)!}{(a+b -1)!}.\]
For $a = |S|+1$ and $b = n-|S|$, we get that the Shapley value of player $i\in N$ is
\begin{multline*}
Sh(i) = \mathlarger{\mathlarger{\sum}}_{S\subseteq N_{-i}} \frac{|S|!\cdot(n-|S|-1)!}{n!}\cdot v'_i(S)
 = \mathlarger{\mathlarger{\sum}}_{S\subseteq N_{-i}} \frac{(a-1)!\cdot(b-1)!}{(a+b-1)!}\cdot v'_i(S)=\\=
\mathlarger{\mathlarger{\sum}}_{S\subseteq N_{-i}} \int \limits_0^1 x^{a-1}(1-x)^{b-1}\cdot v'_i(S) \ dx=
\mathlarger{\mathlarger{\sum}}_{S\subseteq N_{-i}} \int \limits_0^1 x^{|S|}(1-x)^{n-|S|-1}\cdot v'_i(S) \ dx=\\=
\mathlarger{\int \limits_0^1}\mathlarger{\sum}_{S\subseteq N_{-i}} x^{|S|}(1-x)^{n-|S|-1}\cdot v'_i(S) \ dx = 
\mathlarger{\int \limits_0^1}\mathlarger{\sum}_{S\subseteq N_{-i}} x^{|S|}(1-x)^{(n-1)-|S|}\cdot v'_i(S) \ dx.
\end{multline*}
The idea of expressing the Shapley value in the form of this integral originates from \citet{Owen} and is very well described and used for a simulation method for voting games by \citet{Leech2003}. We will refer to formula
\begin{equation}
Sh(i) = \mathlarger{\int \limits_0^1}\sum_{S\subseteq N_{-i}} x^{|S|}(1-x)^{(n-1)-|S|}\cdot v'_i(S) \ dx
\label{formula::integralrep}
\end{equation}
as the \textit{integral representation of the Shapley value}.

\subsection{Expected value representation}
For any $x\in\mathbb{R}$ and finite set $\mathcal{H}$, let us introduce the notation
\begin{note}
\mathbb{P}^{\mathcal{H}}_x(H) = x^{|H|}(1-x)^{|\mathcal{H}|-|H|} \qquad \text{for} \ H\subseteq \mathcal{H}.
\label{note::ph}
\end{note}
where $|H|$ stands for the cardinality (number of elements) of set $H$.
\begin{lemma}
For any $x\in[0, 1]$ and any finite set $\mathcal{H}$, $\mathbb{P}^{\mathcal{H}}_x:2^\mathcal{H}\to [0,1]$ is a probability distribution.
\label{lemma::ph}
\end{lemma}
\begin{proof}
If $0\le x\le1$ then $\mathbb{P}^{\mathcal{H}}_x\ge 0$. If we calculate the sum of the $\mathbb{P}^{\mathcal{H}}_x$ values by grouping the subsets of $\mathcal{H}$ by cardinality, we get that
\begin{multline*}
\sum_{H\subseteq\mathcal{H}} \mathbb{P}^{\mathcal{H}}_x(H) =
\mathlarger{\mathlarger{\sum}}_{k = 0}^{|\mathcal{H}|}\sum_{\substack{H\subseteq\mathcal{H}\\|H| = k}}\mathbb{P}^{\mathcal{H}}_x(H) =
\mathlarger{\mathlarger{\sum}}_{k = 0}^{|\mathcal{H}|}\sum_{\substack{H\subseteq\mathcal{H}\\|H| = k}}x^{|H|}(1-x)^{|\mathcal{H}|-|H|} = \\ =
\mathlarger{\mathlarger{\sum}}_{k = 0}^{|\mathcal{H}|}\sum_{\substack{H\subseteq\mathcal{H}\\|H| = k}}x^{k}(1-x)^{|\mathcal{H}|-k} = 
\sum_{k = 0}^{|\mathcal{H}|}\binom{|\mathcal{H}|}{k}x^{k}(1-x)^{|\mathcal{H}|-k} =
(x + (1-x))^{|\mathcal{H}|} = 1.
\end{multline*}
\end{proof}
\noindent If we combine this notation and the integral representation Formula \eqref{formula::integralrep}, we get that
\begin{equation}
Sh(i) = \mathlarger{\int \limits_0^1}\mathlarger{\sum}_{S\subseteq N_{-i}} x^{|S|}(1-x)^{(n-1)-|S|}\cdot v'_i(S) \ dx = 
\mathlarger{\int \limits_0^1}\mathlarger{\sum}_{S\subseteq N_{-i}} \mathbb{P}^{N_{-i}}_x(S) \cdot v'_i(S) \ dx.
\label{formula::integral_tmp}
\end{equation}
Let us introduce the notation
\begin{note}
\mathbb{P}_x(i, S) = \mathbb{P}^{N_{-i}}_x(S)=x^{|S|}(1-x)^{|N_{-i}|-|S|} = x^{|S|}(1-x)^{(n-1)-|S|}
\end{note}
for each $i\in N$ and $S\subseteq N_{-i}$. Substituting it into Formula \eqref{formula::integral_tmp}, we get that
\[Sh(i) = \mathlarger{\int \limits_0^1}\mathlarger{\sum}_{S\subseteq N_{-i}} \mathbb{P}_x(i, S) \cdot v'_i(S) \ dx.\]
As $S\mapsto\mathbb{P}_x(i, S)$ is a probability distribution (based on Lemma \ref{lemma::ph}), the expression under the integral \[\sum_{S\subseteq N_{-i}} \mathbb{P}_x(i, S) \cdot v'_i(S),\]is, by definition, the expected value of the random variable $v'_i:S\mapsto v(S\cup\{i\})-v(S)$.
Using the notation
\begin{note}
\mathbb{E}_x(v'_i) = \sum_{S\subseteq N_{-i}}\mathbb{P}_x(i, S)\cdot v'_i(S),
\label{note::3}
\end{note}
the Shapley value of player $i$ can be expressed in the concise form
\begin{sh}
Sh(i) = \int \limits_0^1 \mathbb{E}_x(v_i') \ dx.
\label{sh::3}
\end{sh}
\begin{rem}
\label{rempx}
The interpretation of the Shapley value is the average marginal contribution or equivalently the expected marginal contribution of the given player to a random coalition. The difference between Formula \eqref{sh::1} and Formula \eqref{sh::3} is how this coalition is formed. 
\begin{itemize}
\item In case of Formula \eqref{sh::1} all players arrive in a random order to form a chain of growing coalitions from $\emptyset$ to $N$. Payer $i$'s Shapley value is his/her expected marginal contribution to the coalition formed by players who arrived before him/her.
\item Formula \eqref{note::3} is player $i$'s expected marginal contribution to a random coalition formed by simultaneous decisions of all the other players. Imagine that each player except $i$ independently makes a binary decision about joining a cooperating coalition 
with probability $x$. 
$\mathbb{E}_x(v_i')$ is player $i$'s expected marginal contribution to this coalition.
As the integral itself is an expected value, Formula \eqref{sh::3} can be interpreted as player $i$'s expected marginal contribution when the probability of cooperation is also random (with uniform distribution over the interval [0, 1].) 
\end{itemize}
\end{rem}

\section{Linearly representable games}
In this section we introduce the linear representation of games with some examples and basic properties.
\begin{defi}
An ordered pair $\big(f, \{a_j:j\in N\}\big)$ is a linear representation of game $G = (N,v)$ if
\begin{itemize}
\item $a:N\to\mathbb{N}$ is a weighting of the players, and 
\item $f$ is an $f:\mathbb{N}\to\mathbb{N}$ function 
\end{itemize}
 such that the value of each coalition is \[v(S) = f\left(\sum \limits_{j\in S} a_j\right).\]
\label{def::linrep}
\end{defi}
\noindent
If $\big(f, \{a_j:j\in N\}\big)$ is a linear representation of game $G = (N,v)$, we will use the notation 
\begin{note}
a_S = \sum_{j\in S} a_j
\label{note::as}
\end{note}
where $a_j$ denotes the weight of player $j\in N$.\\

\begin{rem}
It is easy to see that every TU game has a linear representation. For example, we can number the players from $0$ to $n-1$ and assign $2^\ell$ to the $\ell^{th}$ player, and let $f(k) = v(S)$ (for $1\le k\le 2^n$), where $S$ is the unique coalition for which $k = a_S$. 
\label{rem::all_lin_rep}
\end{rem}
Remark \ref{rem::all_lin_rep} demonstrates that a linear representation exists for any TU game with at most exponentially large weights, but this representation is not useful, as the point is to make the set $\{a_S \ |\ S\subseteq N\}$ as small as possible, so let us exclude these extreme cases.
\begin{defi}
The linear representation $\big(f, \{a_j:j\in N\}\big)$ is trivial, if the function $S\mapsto a_S$ is injective. 
\end{defi}
In other words a trivial linear representation is just a renaming of the players and the coalitions, which does not help to calculate the Shapley value. The interesting cases are defined as follows.
\begin{defi}
A game $G=(N,v)$ is linearly representable if it has a linear represenation that is not trivial.
\end{defi}
Unfortunately, it is absolutely not clear how to compute a non-trivial linear representation of a given game or how to decide if such representation exists. However, many game classes have a ``natural'' linear representation. Let us consider a few examples.
\begin{ex}[weighted voting games]
Let $w_1,w_2,\dots,w_n \text{ and } q \in \mathbb{N}$ and let the value of a coalition $S\subseteq N$ be as follows. \[v(S) = \begin{cases}1 & \text{if} \ \ \sum \limits_{i \in S} w_i \ge q,\\ 0 & \text{if} \ \ \sum \limits_{i \in S} w_i < q. \end{cases}\]
Weighted voting games are linearly representable, with $a_i=w_i$ and 
\[f(x) = \begin{cases}1 & \text{if} \ \ x \ge q,\\ 0 & \text{if} \ \ x < q. \end{cases}\]
\end{ex}
\begin{ex}[bankruptcy games]
Let $A, l_1,l_2,\dots l_n \in \mathbb{N}$ and let the characteristic function be \[v(S) = \max\left\{0, A - \sum_{j \not \in S } l_j\right\}.\]
Bankruptcy games are also linearly representable. Here $a_i=l_i$ and 
$f(x) = \max\left\{0, A - \left(L-x\right) \right\},$
where $L = \sum l_j$.
\end{ex}
For the interpretation of the characteristic function and more details on voting games and bankruptcy games, see \citet{jatekelm} or \citet{thomson2015axiomatic}. 
\begin{ex}[liability games]
Liability games have recently been introduced by \citet{csoka2019liability} and are similar to bankruptcy games. Let $N=\{0,1,\dots,k\}$ and let $A, l_1,l_2,\dots ,l_k \in \mathbb{N}$  where $A < \sum l_i$, and let the characteristic function be \[v(S) = \begin{cases} \min\{ A, \sum \limits_{j  \in S \setminus \{0\} } l_j\}, & \text{if} \ 0 \in S, \\ \max\{ 0, A- \sum \limits_{j  \notin S \cup \{0\} } l_j\}, & \text{if} \ 0 \notin S. \\\end{cases}\]
Because of the two cases in the characteristic function, it is not completely trivial how to get a linear representation of a liability game. The trick is to assign a sufficiently large number to player 0 (that represents a defaulting firm, which has no liabilities to itself), in a way that we can tell from the sum of the weights weather or not the firm is in the coalition. So let $L=\sum l_i+1$, $a_0=L$, $a_i=l_i$ for $i\ge 1$ and \[f(x)=\begin{cases} \min\{ A, x-L\}, & \text{if} \ x \ge L, \\ \max\{ 0, A- (L-1-x)\}, & \text{if} \ x < L. \\\end{cases}\].  
\end{ex}
The Shapley value for all of these games are NP-hard to compute (see \citet{complex}, \citet{AZIZ2013499} and \citet{CSOKA2021}). 
On the other hand, we know that it is pseudo-polynomial for voting games and bankruptcy games (see \citet{mann1962values} and \citet{AZIZ2013499}) and in this paper we put the icing on the cake and provide a pseudo-polynomial algorithm for every linearly representable game.

\section{Shapley value of a single player}

Let $\big(f, \{a_j:j\in N\}\big)$ be a linear representation of game $G=(N,v)$ and let $i\in N$ be any of the players. We would like to efficiently calculate the expected marginal contribution of player $i$ based on formula \eqref{note::3}:
\[\mathbb{E}_x(v_i') = \sum_{S\subseteq N_{-i}}\mathbb{P}_x(i, S)\cdot v'_i(S).\]

\subsection{Notations and basic properties}
Let us introduce notations
\begin{note}
\mathbb{P}_x(i, k) = \sum\limits_{\substack{S\subseteq N_{-i}\\a_S = k}} \mathbb{P}_x(i, S).
\end{note}
and 
\begin{note}
\mathbb{P}_x(S, k) = \sum_{\substack{H\subseteq S\\ a_H = k}} \mathbb{P}^S_x(H)
\label{note::pxsk}
\end{note}
for $S\subseteq N$ and $k\in\mathbb{Z}$.

\begin{rem}
\label{rem::excepts}
For certain values of $k\in\mathbb{Z}$ (for example when $k<0$ or $k>a_S$, but it can also happen ``accidently") there is no such $H\subseteq S$ that $a_H=k$, so the right hand side of \eqref{note::pxsk} is the empty sum. In such cases, naturally $\mathbb{P}_x(S, k) = 0$. The same convention holds for $\mathbb{P}_x(i, k)$.
\end{rem}

\begin{rem}
Based on Remark \ref{rempx}. $\mathbb{P}_x(i, S)$  is the distribution of a random coalition formed by players except $i$, when they independently decide to cooperate with probability $x$. Following this logic, $\mathbb{P}_x(i, k)$ is the probability that they form a coalition of weight $k$ and in general, $\mathbb{P}_x(S, k)$ is the probability that players in an arbitrary coalition $S$ form a subcoalition of weight $k$.
\end{rem}

It may be worth to point out that, for $k=0$, $\mathbb{P}_x(i, k)$ is not zero but the probability of forming the empty coalition, which is $(1-x)^{n-1}$ and $\mathbb{P}_x(S, 0) = (1-x)^{|S|}$ in general. We summarize the properties of these variables as follows.
\begin{lemma}\hspace{\fill}
\begin{enumerate}[a)]
\item For all $S\subseteq N$ and $x\in[0,1]$, function $k\mapsto\mathbb{P}_x(S, k)$ is a probability distribution over the integers, given by formula \[\mathbb{P}_x(S, k) = \sum\limits_{\substack{H\subseteq S\\a_H = k}} x^{|H|}(1-x)^{|S|-|H|}.\]
\item If $S=\emptyset$ \[\mathbb{P}_x(S, k)=\mathbb{P}_x(\emptyset, k) = \begin{cases} 1 & \text{if} \ \ k = 0 \\ 0 & \text{otherwise.}\end{cases}\]
\item If $S = N_{-i}$ \[\mathbb{P}_x(S, k) = \mathbb{P}_x(N_{-i}, k) = \mathbb{P}_x(i, k).\]
\item If $j\in N$ is a player and $S\subseteq N$ is a coalition such that $j\notin S$ then for each $0\le k \le a_{S\cup\{j\}}$
\begin{equation}
\mathbb{P}_x(S\cup\{j\}, k) = 
\begin{cases}
\ 0 & \text{ if } a_S < k \text{ and } k < a_j,\\
(1-x)\cdot \mathbb{P}_x(S, k) & \text{ if }k\le a_S \text{ and } k<a_j,\\
(1-x)\cdot \mathbb{P}_x(S, k)+x\cdot \mathbb{P}_x(S, k-a_j)&\text{ if }k\le a_S\text{ and } a_j\le k,\\
\hspace{\fill}x\cdot \mathbb{P}_x(S, k-a_j)& \text{ if }a_S < k \text{ and } a_j \le k.
\end{cases}
\label{formula::px_rec}
\end{equation}
\end{enumerate}
\label{lemma::px_rec}
\end{lemma}
\begin{rem}
\begin{sloppypar}
We can combine the four cases of Formula \eqref{formula::px_rec} by applying the convention that \mbox{$\mathbb{P}_x(S, k)=0$} whenever it is an empty sum and put it in a simplified form as follows.
\end{sloppypar}
\begin{equation}
\mathbb{P}_x(S\cup\{j\}, k) = (1-x)\cdot \mathbb{P}_x(S, k)+x\cdot \mathbb{P}_x(S, k-a_j).
\end{equation}
\end{rem}

\begin{proof}
To prove the only non-trivial part, the recursive Formula \eqref{formula::px_rec}, let us consider the family of coalitions
\[\Gamma_{S, k} = \{H\subseteq S \ |\ a_H = k\}\]
for $S\subseteq N$ and $0\le k \le a_S$, i.e.
\[\mathbb{P}_x(S, k) = \sum_{H\in\Gamma_{S, k}} \mathbb{P}^{S}_x(H).\]
In words, $\Gamma_{S, k}$ is a family of sub-coalitions of $S$ with weight $k$, so the coalitions that are considered in the calculation of $\mathbb{P}_x(S, k)$. We can split $\Gamma_{S\cup\{j\}, k}$ into the disjoint union\footnote{Disjoint union is denoted by $\sqcup$.} of subsets
\[\Gamma_{S\cup\{j\}, k} = \underbrace{\{S\in\Gamma_{S\cup\{j\}, k} \ | \ j\notin S\}}_{\Gamma_{S, k}}\sqcup\underbrace{\{S\in\Gamma_{S\cup\{j\}, k} \ | \ j \in S\}}_{\text{let us denote this by }\Gamma^*_{S,j,k}}=\Gamma_{S, k}\sqcup\Gamma^*_{S,j,k},\]
which implies that $\mathbb{P}_x(S\cup\{j\}, k)$ can be split into subsums
\begin{equation}
\label{2terms}
\mathbb{P}_x(S\cup\{j\}, k) = \sum_{H\in\Gamma_{S\cup\{j\}, k}}\mathbb{P}^{S\cup\{j\}}_x(H) = \underbrace{\sum_{H\in\Gamma_{S, k}}\mathbb{P}^{S\cup\{j\}}_x(H)}_* + \underbrace{\sum_{H\in\Gamma^*_{S,j, k}} \mathbb{P}^{S\cup\{j\}}_x(H)}_{**}.
\end{equation}
The first term in Equation \eqref{2terms} is 
\[(*)=\sum_{H\in\Gamma_{S, k}}\mathbb{P}^{S\cup\{j\}}_x(H)=
\sum_{H\in\Gamma_{S, k}}x^{|H|}(1-x)^{|S\cup\{j\}| - |H|}=
(1-x)\underbrace{\sum_{H\in\Gamma_{S, k}}\underbrace{x^{|H|}(1-x)^{|S| - |H|}}_{\mathbb{P}^S_x(H)}}_{\mathbb{P}_x(S,k)}.\]
When $k> a_S$ then $\Gamma_{S, k}=\emptyset$ and $(*)$ is an empty sum. If we omit this term we get the first or the fourth case in Formula \eqref{formula::px_rec}. As for the second term, the mapping
\[\Gamma^*_{S,j,k}\ni H \xleftrightarrow[\quad H'\cup\{j\}\ \reflectbox{\scriptsize$\mapsto$} \ H'\quad]{H\ \mapsto \ H\setminus\{j\}} H'\in\Gamma_{S, k-a_j}\]
is a bijection between $\Gamma^*_{S,j,k}$ and $\Gamma_{S, k-a_j}$, therefore the term $(**)$ in Equation \eqref{2terms} is 
\begin{multline*}
\sum_{H\in\Gamma^*_{S,j, k}} \mathbb{P}^{S\cup\{j\}}_x(H)=
\sum_{H'\in\Gamma_{S,k - a_j}} \mathbb{P}^{S\cup\{j\}}_x(H'\cup\{j\})=
\sum_{H'\in\Gamma_{S,k - a_j}} x^{|H'\cup\{j\}|}(1-x)^{|S\cup\{j\}|-|H'\cup\{j\}|}\\
=\sum_{H'\in\Gamma_{S,k - a_j}} x^{|H'|+1}(1-x)^{|S|+1-(|H'|+1)}=
x\underbrace{\sum_{H'\in\Gamma_{S,k - a_j}} \underbrace{x^{|H'|}(1-x)^{|S|-|H'|)}}_{\mathbb{P}^S_x(H')}}_{\mathbb{P}_x(S, k-a_j)}.
\end{multline*}
When $k<a_j$ then $\Gamma^*_{S,j, k} = \Gamma_{S, k-a_j} = \emptyset$, so this expression is $0$ and if we omit it we get the first or second case of Formula \eqref{formula::px_rec}.
\end{proof}

\begin{rem}
In order to have a solid background for the correctness of the algorithm, \mbox{Lemma \ref{lemma::px_rec}} has an elementary algebraic proof. However the statements are trivial based on the interpretation of $\mathbb{P}_x(S,k)$. This is the probability that players in $S$ form a coalition of weight $k$ when they independently decide to cooperate with probability $x$. Therefore, formula 
\[\mathbb{P}_x(S\cup\{j\}, k) = x\cdot \mathbb{P}_x(S, k-a_j)+(1-x)\cdot \mathbb{P}_x(S, k)\]
is simply the law of total probability as follows.
\begin{multline*}
\mathbb{P}(\text{players in $S\cup\{j\}$ form a coalition of weight $k$}) = \\
= \mathbb{P}\left(\substack{\text{players in $S\cup\{j\}$ form}\\\text{a coalition of weight $k$}} \ \big|\ \text{player $j$ joins}\right)\cdot\mathbb{P}(\text{player $j$ joins})\\
+\mathbb{P}\left(\substack{\text{players in $S\cup\{j\}$ form}\\\text{a coalition of weight $k$}} \ \big|\ \text{player $j$ does not join}\right)\cdot\mathbb{P}(\text{player $j$ does not join})=\\
= \mathbb{P}\left(\substack{\text{players in $S$ form a coalition}\\\text{of weight $k-a_j$}} \ \big|\ \text{\sout{player $j$ joins}}\right)\cdot\mathbb{P}(\text{player $j$ joins})\\
+ \mathbb{P}\left(\substack{\text{players in $S$ form a}\\\text{coalition of weight $k$}} \ \big|\ \text{\sout{player $j$ does not join}}\right)\cdot\mathbb{P}(\text{player $j$ does not join}),
\end{multline*}
and the missing terms correspond to impossible events when players form a coalition with negative weight or less then the weight of one of its members. 
\end{rem}

\begin{cor}
\label{cor::px_rec}
Lemma \ref{lemma::px_rec} suggests the following ``calculation" of $\mathbb{P}_x(S, k)$ in time $O(mK)$ where $m = |S|$ and $K = a_S$:
\begin{itemize}
\item List the members of $S$ in any order: $S = j_1, j_2,\dots ,j_m$ and consider their weights $a_{j_1},\dots,a_{j_m}$ in the same order.
\item Define the growing sequence of coalitions: $S_0 = \emptyset$, $S_1 = \{j_1\}$, $S_2 = \{j_1, j_2\}$,\dots,$S_m = S$.
\item Create a $K$-element array of zeros.
\item Initialize the array to $\mathbb{P}_x(\emptyset, k)$, i.e. set the first element to $1$.
\item Update the array dynamically from $\mathbb{P}_x(S_{j-1}, k)$ to $\mathbb{P}_x(S_j, k)$ for $j = 1..m$ using the recursion
\[\mathbb{P}_x(S_j, k) = \begin{cases}(1-x)\cdot \mathbb{P}_x(S_{j-1}, k) & \text{ if }k<a_j,\\x\cdot \mathbb{P}_x(S_{j-1}, k-a_j)+(1-x)\cdot \mathbb{P}_x(S_{j-1}, k)&\text{ if }k\ge a_j,\end{cases}\]
based on Formula \eqref{formula::px_rec} (because the initial zeros, we only have to take into consideration the second and third case).

\end{itemize}
Algorithm \ref{alg::px_rec} provides a schematic pseudocode.
\begin{alg}
\begin{mdframed}[rightline=false,topline=false,bottomline=false,linewidth=1mm,linecolor=gray]
\label{alg::px_rec}
\vspace*{2mm}
\texttt{INPUT: int a[1:m]}\\[2mm]
\hspace*{0.5cm}\texttt{P[ ] <- new array[0:K]}\\[2mm]
\hspace*{0.5cm}\texttt{for k from 0 to K :}\\[3pt]
\hspace*{1cm}\texttt{P[k] <- } $\begin{cases} \texttt{if k = 0 :}& \texttt{1} \\ \texttt{else :} & \texttt{0} \end{cases}$\\[2pt]
\hspace*{0.5cm}\texttt{for j from 1 to m :}\\[2pt]
\hspace*{1cm}\texttt{for k from K to 0 :}\\[3pt]
\hspace*{1.5cm}\texttt{P[k] <- } $\begin{cases} \texttt{if k < a[j] :} & \texttt{(1-x)*P[k]} \\ \texttt{else :} & \texttt{x*P[k-a[j]]+(1-x)*P[k]} \end{cases}$\\[2pt]
\hspace*{0.5cm}\texttt{return P}
\vspace{3mm}
\caption{recursive calculation of $\mathbb{P}_x(S, k)$ based on the weights of player in $S$.}
\end{mdframed}
\end{alg}

\begin{rem}
It would be intuitive to store all of the $\mathbb{P}_x(S_j, k)$ values in a 2-dimensional array (let us say a matrix) \texttt{R[j][k]} of size $(m+1)\times(K+1)$, initialize its first row to \texttt{[1,0,0,\dots,0]}, fill it row-by-row using the more natural recursion\\[3pt] 
\texttt{R[j][k] := } $\begin{cases} \texttt{if a[j] < k:} & \texttt{(1-x)*R[j-1][k]} \\ \texttt{else:} & \texttt{x*R[j-1][k-a[j]]+(1-x)*R[j-1][k]}, \end{cases}$\\[3pt]
and return its last row. The algorithm calculates the exact same thing, but (to save some space) only stores one row of this matrix and dynamically updates it in each iteration from left to right (as we cannot overwrite \texttt{P[k-a[j]]} before we calculate \texttt{P[k]}).
\end{rem}
\end{cor}
The recursive calculation of $\mathbb{P}_x(i, k)$ suggested by Corollary \ref{cor::px_rec} is the core of the algorithm for the Shapley value. The problem (which is why the word ``calculation" is in quotation marks) is that, as a mathematical object, it is a function of a continuous parameter $x$, for which a calculation is meaningless as computers do not understand the concept of real numbers, real functions and cannot handle a continuum number of parameters. We address this problem in the next section.

\subsection{Algorithm and complexity}
Let $G = (N, v)$ be a TU game, let $i\in N$ be a player and let us turn our attention back to the expected marginal contribution of player $i$:
\[\sum_{S\subseteq N_{-i}}\mathbb{P}_x(i, S)\cdot v'_i(S).\]
Its integral between $0$ and $1$ is the Shapley value of player $i$ in general. However, if $G$ is linearly representable and $\big(f, \{a_j:j\in N\}\big)$ is (one of) its linear representation(s), then this formula can be simplified, because the marginal contribution of any player 
\[v'_i(S) = v(S\cup\{i\}) - v(S) = f(a_{S\cup\{i\}} - f(a_S)=f(a_S+a_i)-f(a_S)\]
depends on the coalition only through its weight. If we group the coalitions by weights, we get that

\begin{multline*}
\mathbb{E}_x(v'_i)=\sum\limits_{S\subseteq N_{-i}}\mathbb{P}_x(i, S)\cdot v'_i(S) = 
\sum\limits_{S\subseteq N_{-i}}\mathbb{P}_x(i, S)\cdot\Big(f(a_S+a_i)-f(a_S)\Big) = \\
=\mathlarger{\sum\limits_{k = 0}^{a_{N_{-i}}}} \sum\limits_{\substack{S\subseteq N_{-i}\\a_S=k}} \mathbb{P}_x(i, S)\cdot \Big(f(a_S+a_i)-f(a_S)\Big)
=\mathlarger{\sum\limits_{k = 0}^K} \sum\limits_{\substack{S\subseteq N_{-i}\\a_S=k}} \mathbb{P}_x(i, S)\cdot \Big(f(k  +a_i)-f(k  )\Big)=\\
=\mathlarger{\sum\limits_{k = 0}^K}\underbrace{\left(\sum\limits_{\substack{S\subseteq N_{-i}\\a_S=k}} \mathbb{P}_x(i, S)\right)}_{\mathbb{P}_x(i, k)}\cdot\underbrace{\Big(f(k+a_i)-f(k)\Big)}_{f_i(k)}=
\sum_{k=0}^K \mathbb{P}_x(i, k)\cdot f_i(k)=h_i(x)
\end{multline*}
where $f_i(k) = f(k+a_i)-f(k)$ and $h_i(x) = \sum_{k=0}^K \mathbb{P}_x(i, k)\cdot f_i(k)=\sum_{k=0}^K f_i(k)\cdot\mathbb{P}_x(i, k)$.
%
%
%
%
%
This means that the Shapley value also depends only on function $\mathbb{P}_x(i, k)$ not on $\mathbb{P}_x(i, S)$ and this observation suggests a straitforward method to calculate or rather to estimate it. Consider formula 
\begin{equation}
\label{shapley5}
\sum_{k=0}^K f_i(k)\cdot\mathbb{P}_x(i, k)
\end{equation} 
as a function of $x$ and calculate its integral numerically. Whenever we need the value of the function for a given $x$, represented as a floating point number, we can calculate it in $O(a_Nn)$ time using the algorithm described in Corollary \ref{cor::px_rec}. Though this algorithm is relatively fast, we must have some concerns about its accuracy (it could be a subject to further research). For example $\mathbb{P}_x(S, 0) = (1-x)^{|S|}$, which might underflow if $S$ is large enough. Nonetheless, this is not exactly the algorithm we are looking for. To improve it, we need a few more observations as follows.

\begin{lemma}
\label{lemma::poly_lemma}
For each $S\subseteq N$ function $x\mapsto\mathbb{P}_x(S, k)$ is a polynomial of $x$, its degree is at most $|S|$, and all of its coefficients 
\begin{itemize}
\item are integer numbers, 
\item have absolute value less then or equal to $3^{|S|}$,
\item can be calculated using only additive integer operations (i.e. addition and subtruction but no multiplication).
\end{itemize}
\end{lemma}
\begin{proof}
All statements are trivial for $S = \emptyset$ and are preserved from $S$ to $S\cup\{j\}$ by \mbox{Formula \ref{formula::px_rec}}.
\end{proof}
Now we can rephrase Corollary \ref{cor::px_rec} as a meaningful statement. By ``calculation'' of $\mathbb{P}_x(S, k)$ we do not mean to calculate its numeric value for a given $x$ (as a floating point number), but to calculate its coefficients as an array of integers. One polynomial can be represented as a list of coefficients and the whole seqence 
\[\mathbb{P}_x(S, k) \qquad 0\le k \le a_S\]
can be represented as a list of polynomials (an array of arrays of integers), so its calculation based on the weights of the linear representation $\{a_j\ |\ j\in S\}$ is a valid computational problem. As Formula \eqref{formula::px_rec} is meaningful for polynomials, the algorithm needs only a slight modification, which however, changes its complexity as stated below.
\begin{thm}
\label{thm::px_comp}
Let $\big(f, \{a_j:j\in N\}\big)$ be a linear representation of game $G=(N,v)$, let $S\subseteq N$ be a coalition of size $m=|S|$ and let us assume that the weights $\{a_j \ | \ j\in S\}$ are stored in an $m$-element array of integers. It is possible to calculate the full sequence of coefficients of the polynomials
\[\mathbb{P}_x(S, k) \qquad 0\le k \le a_S\]
 in time $O(m^3K)$ and space $O(m^2K)$, using only additive integer operations, and store them in space $O(m^2K)$ as an array of arrays of integers where $K = a_S+1$. 
\end{thm}
\begin{proof}
As we have already noted, the solution is to apply Algorithm \ref{alg::px_rec} to polynomials as follows.
\begin{itemize}
\item The array \texttt{P[ ]} does not hold numeric values, but pointers to polynomials.
\item The polynomials have to be created and stored separately, each one as a separate array of integers.
\item In the recursive calculation\\[3pt]
\texttt{P[k] <- } $\begin{cases} \texttt{if k < a[j] :} & \texttt{(1-x)*P[k]} \\ \texttt{else :} & \texttt{x*P[k-a[j]]+(1-x)*P[k]} \end{cases}$\\[2pt]
all operations on the right hand side are carried out on arrays of large numbers.
\end{itemize}
Let us check the complexities of these steps.
\begin{itemize}
\item Array \texttt{P} is initialized with pointers to single-element arrays that represent the constant polynomials. This step has linear complexity. 
\item Based on Lemma \ref{lemma::poly_lemma}, each polynomial can be stored as an array of length at most $m$ that contains $O(m)$-bits integers, which has size $O(m^2)$. Therefore the total memory requirement of all of the polynomials is $O(m^2K)$
\item Finally, the evaluation of the recursive formula \texttt{(1-x)*P[k]} or \texttt{x*P[k-a[j]]+(1-x)*P[k]} is linear in the number of bits as we only need to multiply by polynomial $x$, which is a shift of ceofficients or $\pm 1$, which is also linear and then add $2$ or $3$ polynomials.
\end{itemize}
Consequently the calculation inside the nested loop in Algorithm \ref{alg::px_rec} takes time $O(n^2)$, proportional to the size of the array \texttt{P[k]} and only additions and subtractions are needed and that is exactly what we wanted to proove.
\end{proof}

By applying Theorem \ref{thm::px_comp} for coalition $S=N_{-1}$ we will get the following

\begin{cor}
\label{cor::px_comp}
Let $\big(f, \{a_j:j\in N\}\big)$ be a linear representation of game $G=(N,v)$, let $i\in N$ be a player and let us assume that the weights $\{a_j\ |\ j\neq i\}$ are stored in an $(n-1)$-element array of integers. It is possible to calculate the full sequence of coefficients of the polynomials
\[\mathbb{P}_x(i, k) \qquad 0\le k \le K\]
 in time $O(n^3K)$ and space $O(n^2K)$, using only additive integer operations, and store them in space $O(n^2K)$ as an array of arrays of integers where $K = a_{N_{-i}}$. 
\end{cor}

No we are ready to present our main result.
\begin{thm}
    \label{thm::main_thm}
    Let $\big(f, \{a_j:j\in N\}\big)$ be a linear representation of game $G=(N,v)$ and let $i\in N$ be a player. 
    Let us assume that $a_i\le2^n$ and function $f$ can be evaluated in linear time and space.\\
    The Shapley value of player $i$ can be calculated in $O(n^3K)$ time and $O(n^2K)$ space, where $K=a_{N_{-i}}$.
\end{thm}
\begin{rem}
The exponential upper bound for the weights is justified by Remark \ref{rem::all_lin_rep} and can be relaxed, just like the condition about the complexity of function $f$. The statement holds without these conditions but with some technical modifications. 
\end{rem}
\begin{proof}
The Shapley value of player $i$ is the integral of Formula \eqref{shapley5}
\[\sum_k f_i(k)\cdot\mathbb{P}_x(i, k).\]
First, we compute the polynomials $\mathbb{P}_x(i, k)$, which takes time $O(n^3K)$ and space $O(n^2K)$ based on Corollary \ref{cor::px_comp}, and then we finish the minor steps of the calculation as follows. 
\begin{enumerate}
\item We calculate the sequence of marginal contributions $\{f_i(k)\ | \ 0\le k\le K\}$ and
\item calculate Formula \eqref{shapley5} as a polynomial, which takes $nK$ multiplications and $nK$ additions of $O(n)$-bit numbers.
\item Finally we compute its integral analytically (which is the sum of the coefficients of its primitive function). 
\end{enumerate}
It is easy to see that these three minor steps do not take more than $O(n^3K)$ time and $O(n^2K)$ space, so this is the complexity of the whole calculation.
\end{proof}

\begin{ex}
As an illustration, consider a bankruptcy game with four players, $N = \{a, b, c, d\}$ with liabilities \mbox{$l_a = 2$}, $l_b = 3$, $l_c = 5$, $l_d = 7$ and assets value $A=9$. Let us calculate the Shapley value for player $d$. Table \ref{table::one_shapley_1} shows the recursive calculation of the polynomials $\mathbb{P}_x(S, k)$ for $S_0 = \emptyset$, $S_1 = \{a\}$, $S_2 = \{a, b\}$, and $S_3 = \{a,b,c\}= N_{-d}$ and Table \ref{table::one_shapley_2} shows the same polynomials in the form of sums of monomials. Each polynomial is the sum of $(1-x)$ times the element above it and $x$ times the element $l_j$ columns to the left of it in the previus row, where $a_j$ is the weight of the new player (meaning that elements outside the table are considered $0$). The last row of the table contains polynomials $\mathbb{P}_x(N_{-d}, k) = \mathbb{P}_x(d, k)$, and under the table we indicated the values of $f_d(k) = f(k+l_d)-f(k)=\max(0, k-1)-\max(0, k-8)$. Formula \eqref{shapley5} is the sumproduct of these two rows:
\[x^3-2x^2+x + 2(x^3-2x^2+x)+4(-x^2+x)+6(-x^3+x^2)+7(-x^3+x^2)+7x^3 = -3x^3+3x^2+7x,\]
and the Shapley value of player $d$ is the integral of this polynomial from $0$ to $1$, i.e. the sum of the coefficients of its primitive function, \[-\frac{3}{4}+1+\frac{7}{2}=\frac{15}{4}.\]
\label{ex::one_shapley}
\end{ex}
            
\begin{landscape}
\thispagestyle{empty}
    \begin{table}
        \renewcommand{\arraystretch}{1.4}
        \begin{tabular}{lccccccccccc}
            \specialrule{0.15em}{0em}{1em} 
            $k$&0&1&2&3&4&5&6&7&8&9&10\\
            \hline
            $\mathbb{P}_x(\emptyset, k)$&1\\
            $\mathbb{P}_x(\{a\}, k)$&$1-x$&0&$x$\\
            $\mathbb{P}_x(\{a, b\}, k)$&$\boldsymbol{\color{darkred}(1-x)^2}$&0&$x(1-x)$&$x(1-x)$&0&$\boldsymbol{\color{darkblue}x^2}$\\
            $\mathbb{P}_x(\{a, b, c\}, k)$&$(1-x)^3$&0&$x(1-x)^2$&$x(1-x)^2$&0&$\boldsymbol{\color{darkgray}(1-x)}\cdot\boldsymbol{\color{darkblue}x^2}+\boldsymbol{\color{darkgray}x}\cdot\boldsymbol{\color{darkred}(1-x)^2}$&0&$x^2(1-x)$&$x^2(1-x)$&0&$x^3$\\
            \specialrule{0.15em}{1em}{1em} 
        \end{tabular}
        \caption{Recursive calculation of the polynomials $\mathbb{P}_x(S, k)$.}
        \label{table::one_shapley_1}
    \end{table}
    \begin{table}
        \renewcommand{\arraystretch}{1.4}
        \begin{tabular}{lccccccccccc}
            \specialrule{0.15em}{1em}{1em} 
            $k$&0&1&2&3&4&5&6&7&8&9&10\\
            \hline
            $\mathbb{P}_x(\emptyset, k)$&1\\
            $\mathbb{P}_x(\{a\}, k)$&$-x+1$&0&$x$\\
            $\mathbb{P}_x(\{a, b\}, k)$&$x^2-2x+1$&0&$-x^2+x$&$-x^2+x$&0&$x^2$\\
            $\mathbb{P}_x(\{a, b, c\}, k)$&$-x^3+3x^2-3x+1$&0&$x^3-2x^2+x$&$x^3-2x^2+x$&0&$-x^2+x$&0&$-x^3+x^2$&$-x^3+x^2$&0&$x^3$\\
            \specialrule{0.15em}{1em}{1em} 
            \hline
            $f_d(k)$&0&0&1&2&3&4&5&6&7&7&7\\
            \hline
        \end{tabular}
        \caption{Polynomials of Example \ref{ex::one_shapley} expanded to sum of monomials.}
        \label{table::one_shapley_2}
    \end{table}  
    
\end{landscape}

\section{Shapley value of the whole game}
By Theorem \ref{thm::main_thm} we can calculate the Shapley value of a selected player in $O(n^3K)$ time. To calculate the Shapley value for the entire game in $O(n^4K)$ time, we can run the algorithm for each player one by one. In this section, we discuss how to isolate the most resource intensive part of the computation, which is enough to run only once, and calculate all of the $n$ Shapley values much more faster.
\subsection{Reverse recursion}
The key idea behind the improved algorithm is as follows.
\begin{lemma}
\label{lemma::backward}
The recursive formula for $\mathbb{P}_x(S, k)$ in Lemma \ref{lemma::px_rec} is reversible, meaning that the whole sequence
\[\big(\ \mathbb{P}_x(S, k) \ | \ k\in\mathbb{Z}\ \big)\]
can be calculated based on the whole sequence \[\big(\ \mathbb{P}_x(S\cup\{j\}, k) \ | \ k\in\mathbb{Z}\ \big),\]
where $S\subseteq N$ is a coalition and $j\in N$ is a player such that $j\notin S$.
\end{lemma}
\begin{proof}
Formula \eqref{formula::px_rec},
\[\mathbb{P}_x(S\cup\{j\}, k) = 
\begin{cases}
\ 0 & \text{ if } a_S < k \text{ and } k < a_j,\\
(1-x)\cdot \mathbb{P}_x(S, k) & \text{ if }k\le a_S \text{ and } k<a_j,\\
(1-x)\cdot \mathbb{P}_x(S, k)+x\cdot \mathbb{P}_x(S, k-a_j)&\text{ if }k\le a_S\text{ and } a_j\le k,\\
\hspace{\fill}x\cdot \mathbb{P}_x(S, k-a_j)& \text{ if }a_S < k \text{ and } a_j \le k,
\end{cases}\]
can be ``solved" for $\mathbb{P}_x(S, k)$ as follows:
\begin{equation}
\label{formula::backward}
\mathbb{P}_x(S, k) = \begin{cases}\dfrac{\mathbb{P}_x(S\cup\{j\}, k)}{1-x} & \text{ if }0\le k<a_j,\\[15pt] \dfrac{\mathbb{P}_x(S\cup\{j\}, k)-x\cdot \mathbb{P}_x(S, k-a_j)}{1-x}&\text{ if }a_j\le k\le a_S.\end{cases}
\end{equation}
Though this ``backward" formula seems implicit, in the sense that the solution for $S$ is not solely expressed as a function of $S\cup\{j\}$ but $S$ also turns up on the right hand side, this is not a problem. We can evaluate Formula \eqref{formula::backward} as follows. 
\begin{enumerate}
\item First we can calculate $\mathbb{P}_x(S, k)$ for $0\le k < a_j$ explicitly.
\item We can calculate $\mathbb{P}_x(S, k)$ one by one for $a_j\le k\le a_S$ in increasing order of $k$ (so we calculate $\mathbb{P}_x(S, k-a_j)$ before $\mathbb{P}_x(S, k). $When we need $\mathbb{P}_x(S, k-a_j)$, to calculate $\mathbb{P}_x(S, k)$ we already have it as it is calculated in an earier step.
\item When $a_S<k\le a_{S\cup\{j\}}=a_S + a_j$ then $\mathbb{P}_x(S, k)$ does not appear on the right side, so we cannot express it from the equation, but in this case $\mathbb{P}_x(S, k) = 0$, as an empty sum. 
\end{enumerate}
So in each case, there is a unique solution for $\mathbb{P}_x(S, k)$.
\end{proof}
Formula \eqref{formula::backward} is ambiguous in the same way as Formula \eqref{formula::px_rec}. It can be considred 
\begin{itemize}
\item an identity of functions, meaning that the equation holds for all $x\in [0, 1]$ and 
\item a method to calculate $\mathbb{P}_x(S, k)$ as a polynomial of $x$, meaning to calculate its (integer) coefficients.
\end{itemize}

The first interpretation is clear, except that we have a \textit{division by zero} for $x=1$, when $\mathbb{P}_x(S, k)$ is multiplied by $0$ in Formula \eqref{formula::px_rec}, therefore we can not rearrange the equation to it. However this is not a problem, because for $x=1$ Formula \eqref{formula::px_rec} is simply $\mathbb{P}_1(S\cup\{j\}, k) = \mathbb{P}_1(S, k-a_j)$ therefore $\mathbb{P}_1(S, k)$ can be calculated by formula
\[\mathbb{P}_1(S, k) = \begin{cases}1 & \text{for} \ k = a_S, \\ 0 & \text{otherwise}.\end{cases}\]

For the second interpretation, let $\mathbb{Z}[x]$ denote the set of polynomials with integer coefficients and let us clarify what the division by $(1-x)$ means and how we can calculate it.

\begin{lemma}
\label{lemma::division_by_one_minus_x}
For every $h(x)\in \mathbb{Z}[x]$
\[h(1) = 0 \ \Longleftrightarrow \ \exists g\in \mathbb{Z}[x] \text{ such that } h(x) \equiv (1-x)\cdot g(x),\]
where $\equiv$ means that the two polynomials are identical not only as functions but also as polynomials, meaning that they have the same coefficients when expressed as sums of monomials.
Furthermore, the polynomial $g$ is unique and its coefficients are the cumulative sums of the coefficients of $h$, so they can be computed using additive integer operations.
\end{lemma}
\begin{proof}
We have to solve equation 
\begin{multline*}
\lambda_mx^m+\lambda_{m-1}x^{m-1}+\dots+\lambda_1x+\lambda_0
=(1-x)\left(\mu_{m-1}x^{m-1}+\mu_{m-2}x^{m-2}+\dots+\mu_1x+\mu_0\right)\\
=-\mu_{m-1}x^{m}+(\mu_{m-1}-\mu_{m-2})x^{m-1}+(\mu_{m-2}-\mu_{m-3})x^{m-2}+\dots+(\mu_2-\mu_1)x^2+(\mu_1-\mu_0)x+\mu_0,
\end{multline*}
where $h(x) = \lambda_mx^m+\lambda_{m-1}x^{m-1}+\dots+\lambda_1x+\lambda_0$ and we want to calculate $g(x) = \mu_{m-1}x^{m-1}+\mu_{m-2}x^{m-2}+\dots+\mu_1x+\mu_0$. If we compare the coefficients from right to left we get that
\begin{itemize}
\item $\lambda_0 = \mu_0 \Rightarrow \mu_0 = \lambda_0$,
\item $\lambda_1 = \mu_1 - \mu_0 \Rightarrow \mu_1 = \lambda_1+\mu_0 = \lambda_1+\lambda_0$,
\item $\lambda_2 = \mu_2 - \mu_1 \Rightarrow \mu_2 = \lambda_2+\mu_1 = \lambda_2+ \lambda_1+\lambda_0$,\\
 $\vdots$
\item $\lambda_\ell = \mu_\ell - \mu_{\ell-1} \Rightarrow \mu_\ell = \lambda_\ell+\mu_{\ell-1} = \lambda_\ell+\lambda_{\ell-1}+\dots +\lambda_2+ \lambda_1+\lambda_0$, \\
$\vdots$
\item $\lambda_{m-1} = \mu_{m-1} - \mu_{m-2} \Rightarrow \mu_{m-1} = \lambda_{m-1}+\mu_{m-2} = \lambda_{m-1}+\lambda_{m-2}+\dots +\lambda_2+ \lambda_1+\lambda_0$.
\end{itemize}
The last equation $\lambda_m = -\mu_{m-1}$ is redundant and is equivalent to the existence of the solution:
\[\mu_{m-1} = \lambda_{m-1}+\lambda_{m-2}+\dots +\lambda_2+\lambda_1+\lambda_0 = \sum \lambda_\ell - \lambda_{m} \stackrel{?}{=} - \lambda_{m}\Longleftrightarrow \sum \lambda_\ell = h(1) \stackrel{?}{=} 0.\]
\end{proof}

Based on Lemma \ref{lemma::division_by_one_minus_x}, if $h(x)\in\mathbb{Z}[x]$ a polynomial with integer coefficients, such that $h(1) = 0$ then the division $\dfrac{h(x)}{1-x}$ is meaningful and the ``quotient'' is a polynomial $g(x)\in\mathbb{Z}[x]$.

\begin{cor}
\label{cor::backward_calc}
Let $S\subseteq N$ be a coalition and let $j\in N$ be a player such that $j\notin S$. If we have the whole series of 
\[\mathbb{P}_x(S\cup\{j\}, k) \ \text{for} \ 0\le k\le a_{S\cup\{j\}}\]
represented as an array of arrays of integers, then we can calculate the whole series of
\[\mathbb{P}_x(S, k) \ \text{for} \ 0\le k\le a_S\}\]
represented as an array of arrays of integers, using only additive integer operations in time $O(m^2K)$and space $O(m^2K)$, where $m = |S|+1$ and $K = a_S$.
\end{cor}
\begin{proof}
To calculate the numerator in Formula \eqref{formula::backward} we have to
\begin{itemize}
\item do nothing at all, if $k < a_j$,
\item subtract the $\ell^{th}$ coefficient of $\mathbb{P}_x(S, k-a_j)$ from the $(\ell+1)^{th}$ coefficient of $\mathbb{P}_x(S\cup\{j\}, k)$ for $\ell = 0...|S|$, if $k \ge a_j$. 
\end{itemize}
This is $mK$ additions/subtractions of $O(m)$-bit numbers, which takes $O(m^2K)$ time. To divide the polynomials by $(1-x)$ we have to calculate the cummulative sums of the coefficients (based on Lemma \ref{lemma::division_by_one_minus_x}), which is $m$ operations on $O(m)$-bits integers for each polynomial (i.e. for each $0\le k\le K$), which takes $O(m^2K)$ time as well. 
\end{proof}

\subsection{Algorithm and complexity}
As before, let $\big(f, \{a_j:j\in N\}\big)$ be a linear representation of game $G=(N,v)$. Let us introduce the notation
\begin{note}
\mathbb{P}_x(k) = \mathbb{P}_x(N, k)
\end{note}
which is nothing new, but a shorthand $\mathbb{P}_x(S, k)$, when $S = N$.

\begin{rem}
The interpretation of $\mathbb{P}_x(k)$, as a special case of $\mathbb{P}_x(S, k)$ is the probability of forming a coalition with weight $k$ by all of the players when the probability of cooperation is $x$. It is not particularly useful in itself as no player's Shapley value can be calculated based on it directly. The reason to introduce it is to calculate $\mathbb{P}_x(N_{-i}k)$ using the backward recursion.
\end{rem}

The notation $\mathbb{P}_x(k)$ is umbiguous, as it can refer to the numberic value of $\mathbb{P}_x(k)$ as a function of $x$, and can be considered a polynomial, represented as an array of coefficients. We will consider this latter interpretation and will call the set of polynomials
\[\mathbb{P}_x(k) \qquad (0\le k\le a_N)\]
\textit{base polynomials} of the game. 
Now we can combine Theorem \ref{thm::main_thm} and Corollary \ref{cor::backward_calc} to calculate the Shapley value for the entire game. Algorithm \ref{alg::all_shapley_calc} summarizes the main steps of the calculation. 
\begin{alg}
\begin{mdframed}[rightline=false,topline=false,bottomline=false,linewidth=1mm,linecolor=gray]
INPUT: a linear representation of a TU game $\big(f, \{a_j : j\in N\}\big)$
\begin{enumerate}
\item Calculate the base polynomials $\mathbb{P}_x(k) \ (0\le k\le a_N)$ only once.
\item For $i \in N$
\begin{enumerate}
\item Calculate $\mathbb{P}_x(i, k) \ (0\le k\le a_{N_{-i}})$. 
\item Calculate and integrate Formula \eqref{shapley5} as follows.
\begin{enumerate}
\item Calculate the marginal contributions $\{f_i(k)\ | \ 0\le k\le K\}$ and
\item the coefficients of the polynomial $\sum_k f_i(k)\cdot\mathbb{P}_x(i, k)$.
\item Calculate the integral $\int_0^1 h(x) \ dx$ analyically as the sum of the coefficients of its primitive function.
\end{enumerate}  
\end{enumerate}
\end{enumerate}
\vspace{3mm}
\caption{Calculation of the Shapley value of a linearly representable game.}
\label{alg::all_shapley_calc}
\end{mdframed}
\end{alg}

\newpage
Let us summarize the time and space complexities of these steps as follows.
\begin{itemize}
\item By Theorem \ref{thm::main_thm}, step (1) runs in time $O(n^3K)$ and space $O(n^2K)$. 
\item By Corollary \ref{cor::backward_calc} step $2/a$ runs in time and space $O(n^2K)$. 
\item As for step 2/b, 
\begin{itemize}
\item step (i) takes time and space $O(nK)$ even if the marginal contributions $f_i(k)$ are exponentially large,
\item step (ii) is $nK$ multiplications, which might be too slow (it can take time $\Omega(n^3K$) in the worst case), if the marginal contributions are exponentially large and we use the slowest multiplication method,
\item step (iii), the analytical calculation of the integral of a polynomial of degree at most $n$ (independently of the value of $K$) requires negligible time. 
\end{itemize}
\end{itemize}
It is remarkable (and annoying) that Formula \eqref{formula::backward} saves us the headache of repeating all the calculations over and over again for all players, and still, a trivial-looking step, a bunch of multiplications can slow down the entire algorithm. This is of course no problem if the numbers to multiply are ``small'', so first let us formulate this special case of our final result.  

\begin{thm}
    \label{allThm1}
    Let $\big(f, \{a_j:j\in N\}\big)$ be a linear representation of game $G=(N,v)$ and let us assume that 
\begin{itemize}
\item $a_j\le 2^{|N|}$ and 
\item function $f$ can be calculated in linear time and space and 
\item the coalition values are bounded, i.e. $|v(S)|\le B \quad \forall S\subseteq N$ for some global constant $B\in\mathbb{N}$ that is independent of the game.
\end{itemize}
Then the sequence of the Shapley values of all of the players can be calculated in time $O(n^3K)$ and space $O(n^2K)$, where $K=\sum a_j$.
\end{thm}

\begin{proof}
As we have an upper bound for the coalition values, function $f$ is bounded, $|f(k)|\le B$ for all $k\in\mathbb{N}$, so each multiplication in step (ii) in Algorithm \ref{alg::all_shapley_calc} can be done in time $O(n\log(B)) = O(n)$, and this is enough, as we have $nK$ multiplications for each of the $n$ players. 
\end{proof}

As a generalization of the results of \citet{mann1962values} for voting games, we proved that the calculation of the Shapley value for \textit{one} player has the same complexity as the calculation of the Shapely values for \textit{all} players for a linearly representable game with a bounded characteristic function. In general this result does not hold, but still the calculation of the entire sequence of Shapley values is not $n$-times slower than its calculation for only one player.

Based on the backward formula, Formula \eqref{formula::backward}, it is enough to run the main part of the algorithm once, so the bottleneck, step (ii), is to perform a lot of multiplications on large integer numbers. Though this seems a trivial task, we actually do not know its complexity. The straightforward ``grade-school multiplication'' method has a complexity $O(nm)$, where $n$ and $m$ are the number of bits of the terms. Based on a recent result of \citet{Harvey2021}, the complexity to multiply two $n$-bit numbers is $O(n\log(n))$. This leads to the following theoretical upper bound for the complexity of Algorithm \ref{alg::all_shapley_calc}. 

\begin{thm}
    \label{allThm2}
    Let $\big(f, \{a_j:j\in N\}\big)$ be a linear representation of game $G=(N,v)$ and let us assume that 
\begin{itemize}
\item $a_j\le 2^{|N|}$ and 
\item function $f$ can be calculated in linear time and space.
\end{itemize}
Then the sequence of the Shapley values of all of the players can be calculated in time $O(Kn^3\log(n))$ and space $O(n^2K)$, where $K=\sum a_j$.
\end{thm}

Though the \citet{Harvey2021} multiplication has the best known asymptotic complexity, it might not be practical because of the large constant behind the big-O notation. Providing a galactic algorithm for the calculation of the Shapley value is not really useful, so Theorem \ref{allThm2} only provides a theoretical upper bound for its complexity, but we can do better in real-life applications. A TU game with exponentially large coalition values is unrealistic, but under the much more reasonable assumption that the characteristic function is polynomial in the number of players, we get the same complexity without relying on any nontrivial multiplication method. Let us summarize this final result as follows.

\begin{thm}
    \label{allThm1}
    Let $\big(f, \{a_j:j\in N\}\big)$ be a linear representation of game $G=(N,v)$ and let us assume that 
\begin{itemize}
\item $a_j\le 2^{|N|}$ and 
\item function $f$ can be calculated in linear time and space and 
\item $\exists\alpha\in\mathbb{R}^+$ global constant (independent of the game) such that $|v(S)|\le n^\alpha \quad \forall S\subseteq N$.
\end{itemize}
Then the sequence of the Shapley values of all of the players can be calculated in time $O(Kn^3\log(n))$ and space $O(n^2K)$, where $K=\sum a_j$.
\end{thm}

\begin{ex}
\label{ex::all_shapley}
Table \ref{table::all_shapley_1} illustrates the first step of the algorithm for the same bankruptcy game we have in Example \ref{ex::one_shapley}.
It is exactly the same as Table \ref{table::one_shapley_1}, with one extra row for the fourth player that contains the base polynomials.
Table \ref{table::all_shapley_2} shows the second step. There are four versions of the backward formula depending on the excluded player. These four rows can be used to calculate the four Shapley values.
The last case where the fourth player is excluded gives back the third row of Table \ref{table::all_shapley_1}.
From this point the calculation is the same as it is shown by Table \ref{table::one_shapley_1}.
We calculate the sum of these polynomials weighted by the marginal contributions of the excluded player and integrate it from $0$ to $1$ to get that

\begin{eqnarray*}
    Sh_1 &=& \int \limits_0^1-x^3+4x^2+x \ dx = \left[-\frac{3}{4}x^4+\frac{4}{3}x^3+\frac{1}{2}x^2\right]_0^1= -\frac{3}{4}+\frac{4}{3}+\frac{1}{2}=\frac{13}{12},\\
    Sh_2 &=& \int \limits_0^1-3x^3+4x^2+2x \ dx = \left[-\frac{3}{4}x^4+\frac{4}{3}x^3+x^2\right]_0^1= -\frac{3}{4}+\frac{4}{3}+1=\frac{19}{12},\\
    Sh_3 &=& \int \limits_0^1-3x^3+4x^2+4x \ dx = \left[-\frac{3}{4}x^4+\frac{4}{3}x^3+2x^2\right]_0^1= -\frac{3}{4}+\frac{4}{3}+2=\frac{31}{12}, \\
    Sh_4 &=& \int \limits_0^1-3x^3+3x^2+7x \ dx =  \left[-\frac{3}{4}x^4+x^3+\frac{7}{2}x^2\right]_0^1= -\frac{3}{4}+1+\frac{7}{2}=\frac{15}{4}.
\end{eqnarray*}

\end{ex}

\begin{landscape}
    \begin{table}
        \scalebox{0.63} {
            \renewcommand{\arraystretch}{1.4}
            \begin{tabular}{lcccccccccccccccccc}
                \specialrule{0.15em}{1em}{1em} 
                &0&1&2&3&4&5&6&7&8&9&10&11&12&13&14&15&16&17\\
                \hline
                0&1&0&0&0&0&0&0&0&0&0&0&0&0&0&0&0&0&0\\
                $l_1=2$&$-x+1$&0&$x$&0&0&0&0&0&0&0&0&0&0&0&0&0&0&0\\
                $l_2=3$&$x^2-2x+1$&0&$-x^2+x$&$-x^2+x$&0&$x^2$&0&0&0&0&0&0&0&0&0&0&0&0\\
                $l_3=5$&$-x^3+3x^2-3x+1$&0&$x^3-2x^2+x$&$x^3-2x^2+x$&0&$-x^2+x$&0&$-x^3+x^2$&$-x^3+x^2$&0&$x^3$&0&0&0&0&0&0&0\\
                $l_4=7$&$x^4-4x^3+6x^2-4x+1$&$0$&$-x^4+3x^3-3x^2+x$&$-x^4+3x^3-3x^2+x$&$0$&$x^3-2x^2+x$&$0$&$x^3-2x^2+x$&$x^4-2x^3+x^2$&$x^4-2x^3+x^2$&$-x^3+x^2$&$0$&$-x^3+x^2$&$0$&$-x^4+x^3$&$-x^4+x^3$&$0$&$x^4$\\
                \specialrule{0.15em}{1em}{1em} 
            \end{tabular}
        }
        \caption{Base polynomials for the bankruptcy game in Example \ref{ex::one_shapley}.}
        \label{table::all_shapley_1}
    \end{table}
    
    \begin{table}
        \scalebox{0.6} {
            \renewcommand{\arraystretch}{1.4}
            \begin{tabular}{lcccccccccccccccccc}
                \specialrule{0.15em}{1em}{1em} 
                $k$&0&1&2&3&4&5&6&7&8&9&10&11&12&13&14&15&16&17\\
                $\mathbb{P}_x(k)$&$x^4-4x^3+6x^2-4x+1$&$0$&$-x^4+3x^3-3x^2+x$&$-x^4+3x^3-3x^2+x$&$0$&$x^3-2x^2+x$&$0$&$x^3-2x^2+x$&$x^4-2x^3+x^2$&$x^4-2x^3+x^2$&$-x^3+x^2$&$0$&$-x^3+x^2$&$0$&$-x^4+x^3$&$-x^4+x^3$&$0$&$x^4$\\
                \hdashline
                $l_1 = 2$&$-x^3+3x^2-3x+1$&0&0&$x^3-2x^2+x$&0&$x^3-2x^2+x$&0&$x^3-2x^2+x$&$-x^3+x^2$&0&$-x^3+x^2$&0&$-x^3+x^2$&0&0&$x^3$&0&0\\
                $f(k+2)-f(k)$&0&0&0&0&0&0&0&1&2&2&2&2&2&2&2&2&2&2\\\\
                $v_1'$&\multicolumn{18}{r}{$-3x^3+4x^2+x$}\\
                \hdashline
                $l_2=3$&$-x^3+3x^2-3x+1$&0&$x^3-2x^2+x$&0&0&$x^3-2x^2+x$&0&$-x^2+x$&0&$-x^3+x^2$&0&0&$-x^3+x^2$&0&$x^3$&0&0&0\\
                $f(k+3)-f(k)$&0&0&0&0&0&0&1&2&3&3&3&3&3&3&3&3&3&3\\\\
                $v_2'$&\multicolumn{18}{r}{$-3x^3+4x^2+2x$}\\
                \hdashline
                $l_3=5$&$-x^3+3x^2-3x+1$&0&$x^3-2x^2+x$&$x^3-2x^2+x$&0&$-x^3+x^2$&0&$x^3-2x^2+x$&0&$-x^3+x^2$&$-x^3+x^2$&0&$x^3$&0&0&0&0&0\\
                $f(k+5)-f(k)$&0&0&0&0&1&2&3&4&5&5&5&5&5&5&5&5&5&5\\\\
                $v_3'$&\multicolumn{18}{r}{$-3x^3+4x^2+4x$}\\
                \hdashline
                $l_3=7$&$-x^3+3x^2-3x+1$&0&$x^3-2x^2+x$&$x^3-2x^2+x$&0&$-x^2+x$&0&$-x^3+x^2$&$-x^3+x^2$&0&$x^3$&0&0&0&0&0&0&0\\
                $f(k+7)-f(k)$&0&0&1&2&3&4&5&6&7&7&7&7&7&7&7&7&7&7\\\\
                $v_4'$&\multicolumn{18}{r}{$-3x^3 +3x^2+7x$}\\
                \specialrule{0.15em}{1em}{1em} 
            \end{tabular}
        }
        \caption{Calculation of the backward formula.}
        \label{table::all_shapley_2}
    \end{table}
\end{landscape}

\section{Summary and possible extensions}
So far, we have provided a pseudo-polynomial algorithm for the exact calculation of the Shapley value of a linearly representable game. As it turned out, the critical par is the calculation of a data set, the coefficients of the \textit{base polynomials}, which is technically a $K\times n$ matrix, and then the entire sequence of Shapley values can be calculated very quickly, almost as quickly as the Shapley value of a single player. Similar results appeared in the literature for special game types (voting games by \citet{mann1962values}, bankruptcy games by \citet{AZIZ2013499}), but the unified discussion of such game classes is a novelty. In this final section we briefly mention a few extensions, applications, modifications and limitations of the algorithm.

\subsection{Optimizing the algorithm}
The algorithm in its current form has a drawback. The very first step is to create a huge nullmatrix. This makes the algorithm slow if $K$ is large but there are only a polynomial number of nonzero polynomials $\mathbb{P}_x(S, k)$. We can fix this issue, if we only store the nonzero polynomials $\mathbb{P}_x(S, k)$ in a so-called associative array. An associative array is a data structure that stores key-value pairs such that we can access the values ``quickly'' by their unique keys. This way, we do not have to initialize a large matrix of zeros, only a very small data structure that initially stores nothing but the polynomial $\mathbb{P}_x(\emptyset, k) = 1$, and then add only the nonzero new values when the algorithm calculates them.

In the second step of the algorithm we iterate the polynomials $\mathbb{P}_x(S, k)$ in increasing order to correctly calculate the backward formula, we have to keep the data structure sorted by keys, or we have to sort it at the end of the first step. This has some extra computational cost, which is only worth it, if we have a lot of zero polynomials $\mathbb{P}_x(S, k)$. On the other hand, this modification makes the algorithm as fast as possible, by automatically skipping all meaningless operations with unnecessarily stored zeros without any futher information requirement (we do not have to know anything about the number of nonzero probabilities in advance).

\subsection{Application for other game types}

Let us consider airport cost games introduced by \citet{littlechild1973simple}. Each player is asssociated with a cost $c_i \in \mathbb{N}$ and the value of a coalition is 
\[v(S) = \max\{c_i \ | \ i \in S\}.\]

This is not a linear representation of the game, because the value of a coalition is not a function of the sum of the weights of its members, but their maximum. Mathematical intuition suggests that this nonlinearity makes the problem much more complicated, but in fact, not just that we can use a slightly modified version of our algorithm, but it is actually polynomial in this case. To see this, let us first observe that $\mathbb{P}_x(S\cup \{j\}, k) = 0$ unless $k$ is one of the weights of the players, so we only need to calculate the polynomials $\mathbb{P}_x(S\cup \{j\}, c_k)$. It is easy to see that the recursive formula, Formula \eqref{formula::px_rec} for this game class is

\[\mathbb{P}_x(S\cup \{j\}, c_k) =   \begin{cases}
                                \mathbb{P}_x(S, c_k) & \ \text{if} \ \ c_j < c_k, \\
                                \mathbb{P}_x(S, c_k) + (1-\mathbb{P}_x(S, c_k))\cdot x = (1-x)\cdot \mathbb{P}_x(S, c_k) + x  & \ \text{if} \ \ c_j = c_k, \\
                                (1-x)\cdot \mathbb{P}_x(S, c_k)      & \ \text{if} \ \ c_j > c_k.\\
                \end{cases}\] 

If $c_s \le c_j \ \forall s\in S$ then in the first case, when $c_j<c_k$ then of course $\mathbb{P}_x(S\cup \{j\}, c_k) = \mathbb{P}_x(S, c_k) = 0$, which makes the recursion even simpler. This is all we need for the algorithm we discussed in Section 4. We can calculate the $\mathbb{P}_x(S, c_k)$ polynomials recursively, but only $n$ of them are nonzero, so the algorithm to calculate one Shapley value has complexity $O(n^4)$. 
Furtunately, even the recursive calculation is unnecessary as it is easy to see that we can express $\mathbb{P}_x(S, c_k)$ explicitely as follows.

\[\mathbb{P}_x(S, c_k) = \Big(1-(1-x)^{n_k}\Big)\cdot(1-x)^{n_{k+}},\]
where $n_k$ is the number of players with cost exactly $k$ and $n_{k+}$ is the number of players with cost strictly bigger than $k$.
This practically leads to an algorithm to calculate and store binomial coefficients with alternating signes.

\subsection{Limitations}
In 1978 \citet{dori1} introduced a game for sharing the cost of building a network (for example electricity) with one root node and $n$ users. The cost of each coalition is the total cost of egdes they need to supply the entire coalition. The Shapley value of this game is very intuitive: the cost of each edge is paid equally by those players that use it. Airport games that we discussed in the previous subsection are a special case of these cost sharing games with network that consists of only one path. It is remarkable, that a slightly modified version of our new algorithm can calculate the Shapley value of airport games, but it is absolutely not clear how to even represent the cost sharing games in general such that we can apply the algorithm to it. Of course, we know for decades that the Shapley value is a fairly easy combinatorical problem and therefor polynomial for voting games (see. \citet{littlechild1973simple}), but our new algorithm provides a new proof and a new method, while it can do nothing with a cost sharing game in general, for which the Shapley value is also polynomial to calculate. Consequently, it might still have a lot of potential applications and limitations for further research.

\newpage
\bibliography{references2}
\bibliographystyle{plainnat}
\end{document}